\documentclass[runningheads]{llncs}
\usepackage[T1]{fontenc}
\usepackage{graphicx}
\usepackage{makecell}
\usepackage{amsmath}
\usepackage[T1]{fontenc}
\usepackage{booktabs}
\usepackage{multirow}
\usepackage[utf8]{inputenc}

\begin{document}
\title{Classification of User Satisfaction in HRI with Social Signals in the Wild}
%
%
\author{Michael Schiffmann\inst{1}\orcidID{0000-0002-7328-9859},\\Sabina Jeschke \inst{2}\orcidID{0000-0003-2326-8264}\\ \and
Anja Richert\inst{1}\orcidID{0000-0002-3940-3136}}
%
%
\institute{TH Köln - University of Applied Sciences,\\ 
Cologne Cobots Lab, Betzdorfer Str. 2, 50679, Germany \\
\email{\{michael.schiffmann\}\{anja.richert\}@th-koeln.de} \inst{1} \\ 
KI Park e.V. \& FAU - Friedrich-Alexander University of Erlangen-Nuremberg  \inst{2}}

\maketitle              
\begin{abstract} 
Socially interactive agents (SIAs) are being used in various scenarios and are nearing productive deployment. Evaluating user satisfaction with SIAs' performance is a key factor in designing the interaction between the user and SIA. 
Currently, subjective user satisfaction is primarily assessed manually through questionnaires or indirectly via system metrics.  This study examines the automatic classification of user satisfaction through analysis of social signals, aiming to enhance both manual and autonomous evaluation methods for SIAs.
During a field trial at the Deutsches Museum Bonn, a Furhat Robotics head was employed as a service and information hub, collecting an "in-the-wild" dataset. This dataset comprises 46 single-user interactions, including questionnaire responses and video data. 
Our method focuses on automatically classifying user satisfaction based on time series classification. We use time series of social signal metrics derived from the body pose, time series of facial expressions, and physical distance. This study compares three feature engineering approaches on different machine learning models.  
The results confirm the method's effectiveness in reliably identifying interactions with low user satisfaction without the need for manually annotated datasets. This approach offers significant potential for enhancing SIA performance and user experience through automated feedback mechanisms.

\end{abstract}
\section{Introduction}
Socially interactive agents (SIA) are applied in various scenarios, such as in public spaces like a museum \cite{cantucci2022autonomous,List,willeke2001history}.
SIAs can be viewed from a technological and application-specific perspective: On the one hand, it is crucial by definition that they continuously adapt to users. On the other hand, operators and researchers are interested in whether SIAs deliver the desired performance in productive use. For both approaches, the automatic measurement of user satisfaction can be crucial for the evaluation and improvement of SIAs. \\
The integration of UX data is an essential development factor \cite{shourmasti_user_2021,alenljung_user_2017}, particularly in service and information environments where the goal is to assist users.
There are various ways to evaluate SIAs, for instance, indirectly through objective system metrics regarding the efficiency of the SIA’s speech dialogue system, such as the number of turns or dialogue duration  \cite{canizares_automating_2022,walker_paradise_1997}. Alternatively, metrics like the number of different dialogue paths can be used to assess the complexity of potential dialogues, which, according to  \cite{canizares_automating_2022}, is an indicator of efficiency and satisfaction.\\
Direct evaluation of subjective user satisfaction is only possible through questionnaires or interviews.
An approach that captures this implicitly is necessary for the automatic collection of user satisfaction since active feedback requests interrupt the interaction, and post-interaction feedback requests risk low participation. A video-based possibility involves capturing social signals that allow insights into the user's internal state. Nonverbal social signals or cues can be used to make these inferences \cite{muller_detecting_2018} and like \cite{wei_multimodal_2021,schiffmann_predicting_2025,jokinen_modelling_2015} showed that annotated video data can be used to classify quality features like user satisfaction or user experience. Assessing the user’s state is not only of interest for adaptation but also for evaluating the SIA’s performance in specific applications.\\
In this paper, we explore how individual user satisfaction ratings can be automatically classified using social signals as a source for machine learning feature extraction, specifically body language, facial expressions, and distance, to predict user satisfaction. We use an in-the-wild dataset consisting of single user interactions and corresponding post user satisfaction ratings of each particular user. To achieve this, we utilize data collected from a field study with an autonomous SIA deployed as a service and information hub at the Deutsches Museum Bonn. 
Our method distinguishes itself from the state-of-the-art by using an holistic approach to capture social signals from body language, not requiring additional annotations from third parties or users. 
We compare three feature engineering methods to demonstrate how to classify interactions with low user satisfaction and interactions with medium to high user satisfaction ratings. Despite unpredictable conditions, realistic interaction data offer valuable insights, allowing for the study of unbiased interactions with an autonomous SIA. 
In the following, we first discuss the psychological background of social signals and related work. We then present the implemented system and experimental setting. In the methods section, we explain data collection, feature engineering, and the composition of the classification task before we finally present the results and discuss them in terms of the peculiarities of the "in-the-wild" approach and further research avenues arising from this work. 
\section{Related Work}
In social interactions, individuals exchange information through verbal and non-verbal communication with their conversation partners. Social signals, including body language, gestures, and facial expressions, serve various purposes in this context and help convey one's emotional state or stance towards something (social attitude) \cite{poggi_cognitive_2010}.
In HRI, social signals have been effectively utilized to evaluate various psychological constructs, such as rapport \cite{muller_detecting_2018,wang_rapport_2009}, user satisfaction \cite{wei_multimodal_2021,schiffmann_predicting_2025}, user experience \cite{jokinen_modelling_2015}, user engagement \cite{jokinen_modelling_2015}, and affective assignment \cite{mccoll_determining_2014,hong_multimodal_2021} within Russell's circumplex model of affect (two-dimensional model for emotions) \cite{russell_circumplex_1980}.  
Müller et al. \cite{muller_detecting_2018} predicted perceived low rapport versus medium/high rapport in human-to-human group interactions using automatically derived features from social signals. They found that facial cues, including a variety of features such as indicators of happiness, synchrony indicators, the amount of mutual facing, among others were the strongest predictors of low rapport, achieving an average accuracy of 70\%.  
Jokinen and Wilcock \cite{jokinen_modelling_2015} showed in a laboratory HRI Experiment that certain aspects of user experience have a significant correlation with the user behavior and that it is possible to predict the user's experience overall and its aspects of usability, expressiveness, responsiveness, and interface from a manually annotated dataset.
Wei et al. \cite{wei_multimodal_2021} studied user satisfaction (well-coordinated, awkward, unfriendly) in a Wizard-of-Oz interaction with a digital agent, assessing it through user questionnaires and self-assessments. They successfully used deep learning on multi-modal verbal and non-verbal social signals, finding that multi-modal models outperformed uni-modal ones in predicting user satisfaction. 
Schiffmann et al. \cite{schiffmann_predicting_2025} showed that user satisfaction can be predicted based on annotated valence and arousal time series features and the user satisfaction value of the respective users of human-robot interaction which was recorded in the wild.
Using a generated skeleton model, McColl et al. \cite{mccoll_determining_2014} demonstrated how to predict valence and arousal levels by translating the user's body language into angles, velocities, and expansion. This approach was subsequently utilized with a rater-based method for annotating interaction data, resulting in the development of separate valence and arousal models for training and prediction.
However, all these approaches share the commonality of being based on controlled laboratory studies, with training data generated through an annotation process, except in the case of \cite{muller_detecting_2018,schiffmann_predicting_2025}.

\section{System and Setting}
\begin{figure}
    \centering
\includegraphics[width=0.75\linewidth]{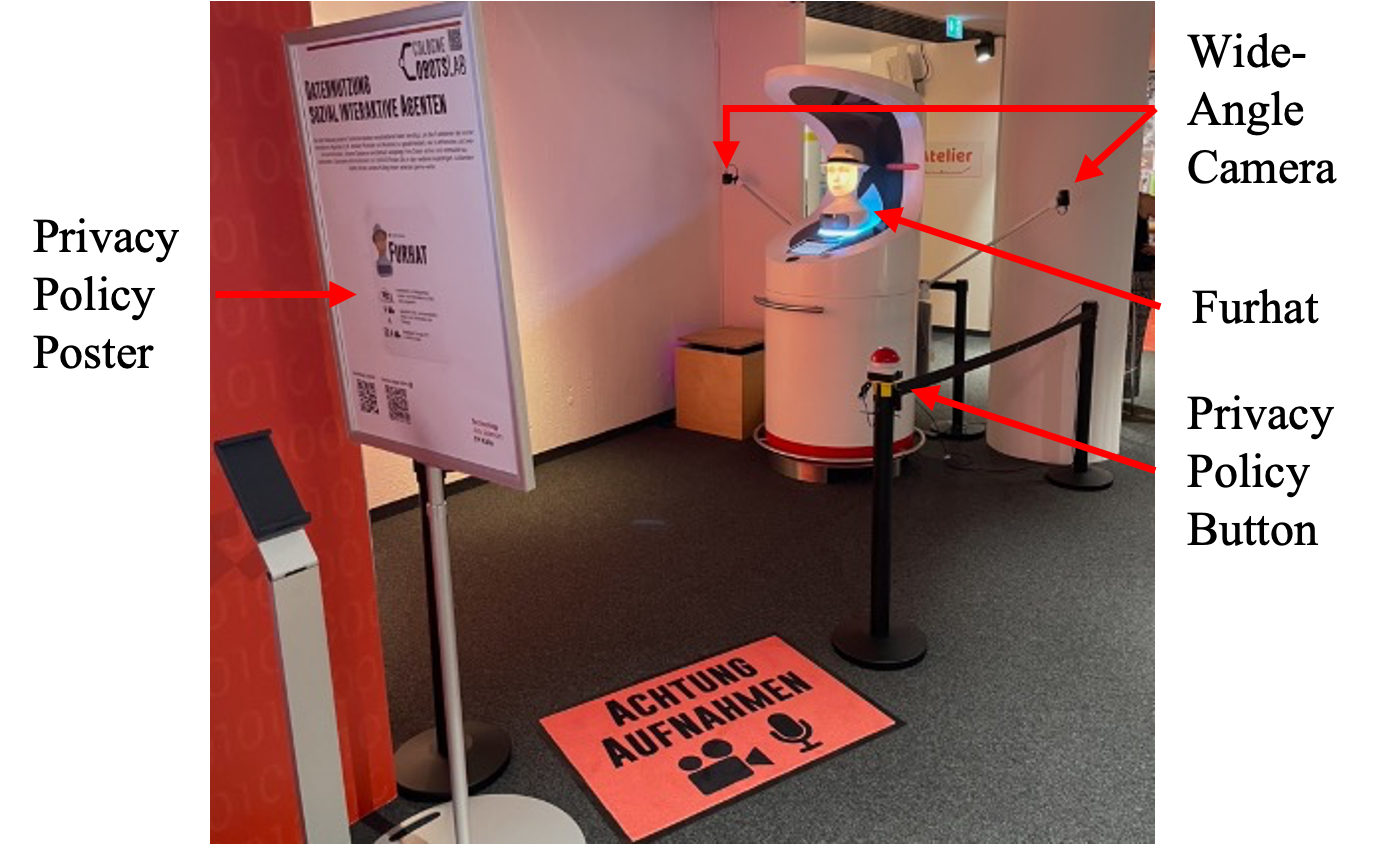}
\caption{Setting of the Field Test in the Museum.} \label{SettingLabeled}
\end{figure}

The field trial took place from July 23, 2024, to August 15, 2024, at the Deutsches Museum Bonn (DMB). The DMB is presenting an exhibition on artificial intelligence with interactive exhibits to bring artificial intelligence closer to a broad audience. The robot "Mira", developed to answer visitor questions, was positioned near the entrance in a designated interaction area. Figure \ref{SettingLabeled} illustrates and labels the experimental setup.

The Furhat robot \cite{furhat} from Furhat Robotics AB is mounted on a special casing with an additional 10-inch tablet displaying the transcribed dialogue during interactions. Two wide-angle cameras are mounted on the sides of Furhat to capture the user fully, as the geometry of the housing prevents full-body capture with the integrated camera alone.
Before interacting with the robotic system, museum visitors must agree to the data processing policy by pressing the privacy policy button, as informed by a poster. Video, audio data, and interaction logs were recorded for research. The Furhat robot remains in idle mode with a blue LED until consent is given. If permission isn't granted and the user is detected by the person management, the robot prompts visitors. Upon agreement, the robot initiates interaction when a user enters the area, with the LED changing to green for listening and red for processing and speaking.
On a technical level, the system is divided into front-end and back-end. The front-end consists of the robot head, which runs a so-called Furhat Skill that uses the Furhat SDK for person management, speech-to-text, and text-to-speech functionalities. The back-end handles session and dialogue management and persistence of interaction data with a unique session ID. 
The system's conversational knowledge is either managed through intent-based natural language understanding with a Q\&A database or, if that fails, generated by a locally executed LLM, Llama 3, as a fallback.
The LLM receives the previous conversation history and the prompt: "This is a conversation between a visitor of the Deutsches Museum in Bonn and a robot named Mira. Mira is programmed to respond briefly and precisely, in no more than two sentences." Additionally, the back-end controls the media recordings and captures material from all cameras. The wide-angle cameras support Furhat's person management by capturing users who are too tall or short for the integrated camera or those who turn aside for too long, preventing premature interaction termination. By implementing an additional person detection with YOLO (yolov5 \cite{jocherUltralyticsYolov5V312020}), a double-check was realized to verify the presence of individuals in the interaction area.

\section{Method}
Our approach to investigate whether the user's social signals can be used to classify user satisfaction rating from time series features automatically is methodologically composed of conducting the experiment itself for data collection, data preparation and preprocessing for the respective feature engineering methods for machine learning.

\subsection{Field Test Execution and Questionnaire}
Museum visitors interacted freely with the robot and were informed solely through a privacy policy poster. After interacting, a researcher invited them to fill out a questionnaire. Due to the general reluctance to complete lengthy questionnaires in the field, we use a custom-designed, concise questionnaire.\\
The questionnaire assessed user satisfaction using five items on a 5-point Likert scale (1 = strongly disagree and 5 = strongly agree). Additionally, demographic data of the users was collected. 
The user is first asked for a statement on "Overall, I am satisfied with the system." which is derived from the first question of the After Scenario Questionnaire \cite{lewis_ibm_1995} and is intended to capture the individual's satisfaction according to their judgment. The second statement is "If I meet this agent again, I would talk to him again." which is supposed to assess the willingness to interact with the agent again; we expect that a satisfactory interaction will result in a desire to re-interact. 
The subsequent three statements target interaction dynamics and focus on the participant's perception of the conversation flow and speaker changes, as these elements are crucial for a seamless interaction. The third statement is "The agent motivated me to continue the conversation" which seeks to evaluate the persuasive ability of the agent, thereby indicating the degree to which the interaction was engaging. The fourth statement, "The alternating listening and speaking was intuitive" addresses the fluidity and naturalness of turn-taking during the interaction. Finally, "It was understandable to me when the agent was listening to me" gauges how clearly the participants could distinguish when the agent was attentive to their input. We use Cronbach's Alpha to check items' internal consistency and interrelation in this custom, not validated scale, ensuring it reliably measures user satisfaction. For the machine learning task, we combine the five items via averaging to a user satisfaction score, which is used as target label. 

\subsection{User Satisfaction Classification}
\begin{figure}
\includegraphics[width=\textwidth]{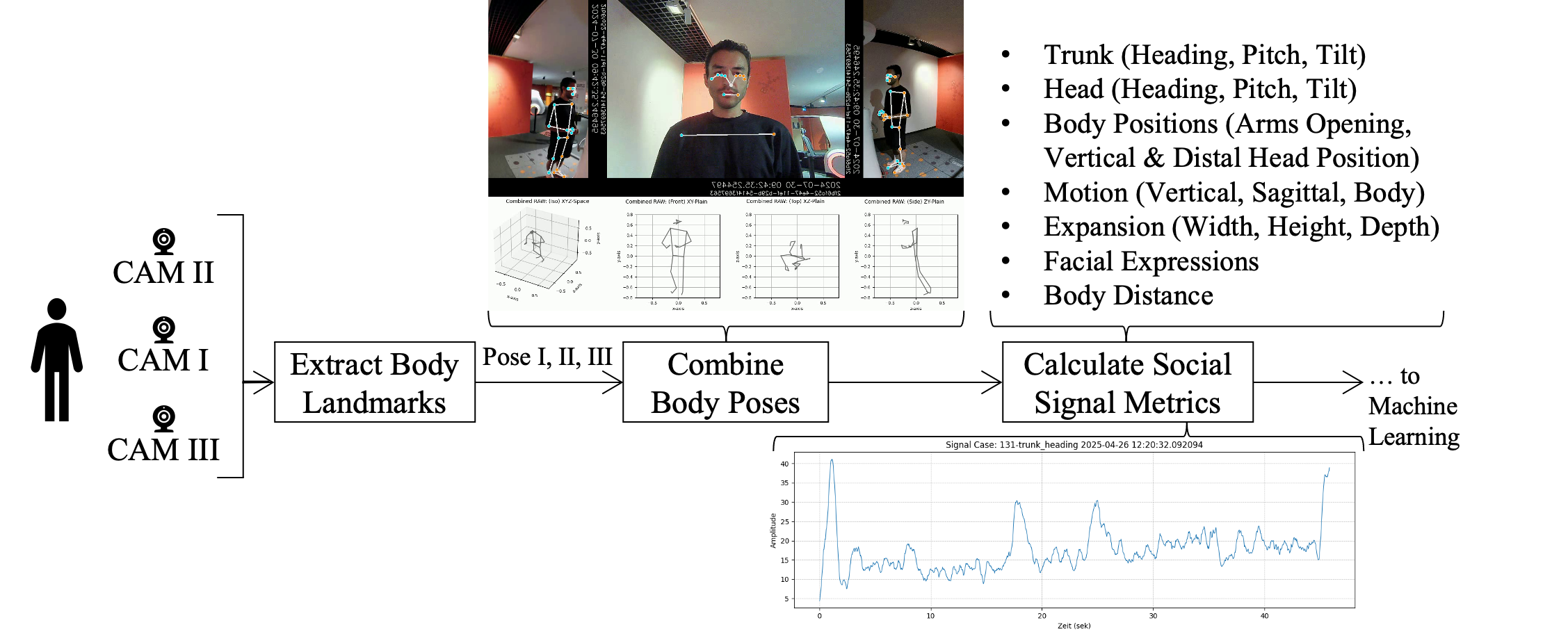}
\caption{Processing Flow to Extract the Body Landmarks from each Camera Frame and Combine them into one Body Posture to Calculate the Different Social Signal Metrics} \label{rapportPipeline}
\end{figure}

Our approach involves the following steps: extracting a body pose, calculating various social signal metrics per camera frame, computing time series-specific features for three different feature engineering approaches, and selecting relevant features for the final training of models. We utilize two approaches for automatic feature extraction, and our own approach with handcrafted features is designed to mimic the coding used in qualitative video analyses. \\
\textbf{Extraction of the Social Signals}
In this study, we use body language, e.g., posture, facial expressions, and the user's distance to the agent. The extraction of body language is realized using MediaPipe \cite{mediaPipe}. A total of 33 body landmarks are extracted for the key nodes on the human body and are returned as normalized X, Y, Z coordinates along with a visibility value.
Figure \ref{rapportPipeline} shows the workflow from capturing the person to outputting the metrics. The landmarks generated from the three camera perspectives are synchronized to 30 frames per second and then fused in the combining step, where visibility and distance between the landmarks are considered, resulting in one body pose. Initially, the poses are rotated because the cameras were not aligned, yet the landmark detector assumes strict vertical and horizontal alignment of the cameras. 
The facial expressions are extracted with py-feat \cite{cheong_py-feat_2023}, which uses the MediaPipe face mesh with 468 3D landmarks  \cite{mediaPipe}. The output of py-feat predicts seven expressions: happiness, anger, disgust, fear, sadness, surprise, and neutral. For the distance measure, we use the approach by \cite{noauthor_mediapipe_nodate} and utilize the diameter of the user's iris to estimate the distance to the front camera since this can be assumed to have a fixed size. Body language, such as nodding, facing towards, and leaning in or out, encodes attitudes towards the SIA through posture over time. Our concept for capturing these social signals adapts the approach of McColl et. al. \cite{mccoll_determining_2014} by translating a user's body posture into angle and velocity metrics such as head and torso orientation (heading, pitch, tilt), body dimensions and its velocity in three spatial directions, head position (vertical \& sagittal) and arm opening —resulting in 16 time-series metrics including distance (see online Appendix \cite{michael_icsr2025} for more details). We use this implicit holistic approach to capture body language because we have found that users show few explicit gestures \cite{schiffmann_evaluation_2024} such as affirmative hand gestures, nodding, head shaking or expressive facial expressions as commonly used in annotation-based research cf. \cite{jokinen_modelling_2015,wei_multimodal_2021}. Figure \ref{rapportPipeline} shows the resulting time series for head heading (orientation to the agent) over the interaction period. \\
\textbf{Data generation and Preprocessing}
For the final dataset, only single interactions were selected where the participants were mostly alone in the interaction area. Additionally, only interactions linked to a fully completed questionnaire were chosen. The selection rules lead to a reduced data set, but this prevents the time series from containing disturbances, e.g. because another person has entered the interaction area or the person has changed, which leads to a distorted data set. We use only the main phase of the interaction because while the user is entering or leaving the interaction area, the body pose can only be combined in insufficient quality, leading to highly distorted social signals. We observed occasional body pose misclassifications during the interaction, leading to distortions and missing values, which we corrected by applying linear interpolation and smoothing techniques on the time series.\\
\textbf{Feature Engineering Approaches}  
The extraction of social signals results in 23 time series to generate social signal metrics, where 16 metrics are generated from the body and seven from the face. 
Analyzing and generating features of these time series is challenging as they must comprehensively represent the connection between users state and user satisfaction \cite{fulcher_feature-based_2017}. We compare three state-of-the-art feature engineering approaches to determine their effectiveness in classifying user satisfaction.\\
\textbf{Feature Engineering with tsfresh}
The tsfresh library is an extensive Python tool for automatically extracting various features from time series data (used version 0.21.0) \cite{christ_time_2018}. It enables precise analysis and classification by generating numerous relevant features and selecting suitable attributes for machine learning. \\
\textbf{Feature Engineering with catch22}
The catch22 library follows the idea to calculate 22 of the most important time series features, such as linear and nonlinear autocorrelation, successive differences, value distributions, and outliers (used version pycatch22 0.4.5) \cite{lubba_catch22_2019}. These features have been shown to capture key characteristics and patterns to facilitate effective time series classification and analysis. \\
\textbf{Handcrafted Feature Engineering}
We employ an approach mimicking interaction annotation by establishing angular ranges for each social signal metric which is crucial for interaction. For instance, for the head, we define areas that distinguish between "looking at" and "looking away" (see online Appendix \cite{michael_icsr2025} for more details). The social signal time series is assigned to these zones by calculating the average over a 0.5-second time window to determine the appropriate zone. Based on these assignments per time window, we compute various features per metric: average, minimum, maximum, standard deviation, the most frequently occurring zone, the distribution of individual zones, the number of zone transitions, and the longest and shortest durations within a zone. Features based on the time series' value, such as mean, minimum, maximum, and standard deviation, are also calculated for further classification. This approach uses the time series of the trunk and head, body speed, distance, and facial expressions (confidence threshold is used) for a simple and understandable assignment of zones.\\
\textbf{Model and Training}
Each of the three feature engineering methods produce a large number of features, therefore we use the "SelectKBest" method available in scikit-learn \cite{buitinck_api_2013}, which selects the best k features based on ANOVA F-value analysis.
Due to the small dataset size of N=46 instances, we simplify the classification task by dividing the dataset into two classes: the bottom 33\% percentile represents the class of dissatisfied users. In contrast, the upper 66\% represents moderately to highly satisfied users. The simplifications helps to prevent overfitting since models need less data per class. 
The classification task remains relevant, as there is a pressing need for improvement of the robot, primarily when the user is dissatisfied. In a preliminary test with the dataset, we explored artificially augmenting the data by employing SMOTE \cite{chawla_smote_2002} and duplicating the dataset through vertical mirroring; however, we discarded these approaches due to the lack of observed benefits.
The selected ten best features were standardized and used for comparison to train a diverse set of algorithms  Random Forest (ensemble method), Support Vector Machine and Logistic Regression (linear models), and Naive Bayes Classifier model (probalistic model) to compare them against each other. 
We use the leave-one-out cross-validation method (LOOCV), since our dataset consists of 46 interactions we aim to prevent overfitting. We complemented this with a parameter search to find optimal parameters for the different models. 
We aim to demonstrate the feasibility of this automatic approach, which works with labeled datasets instead of manually annotated data. Therefore, the performance of the models is evaluated by analyzing accuracy and assessing the model's discriminative ability, F1 score, precision, ROC-AUC, and recall. 

\section{Results}
\begin{figure}
\includegraphics[width=\textwidth]{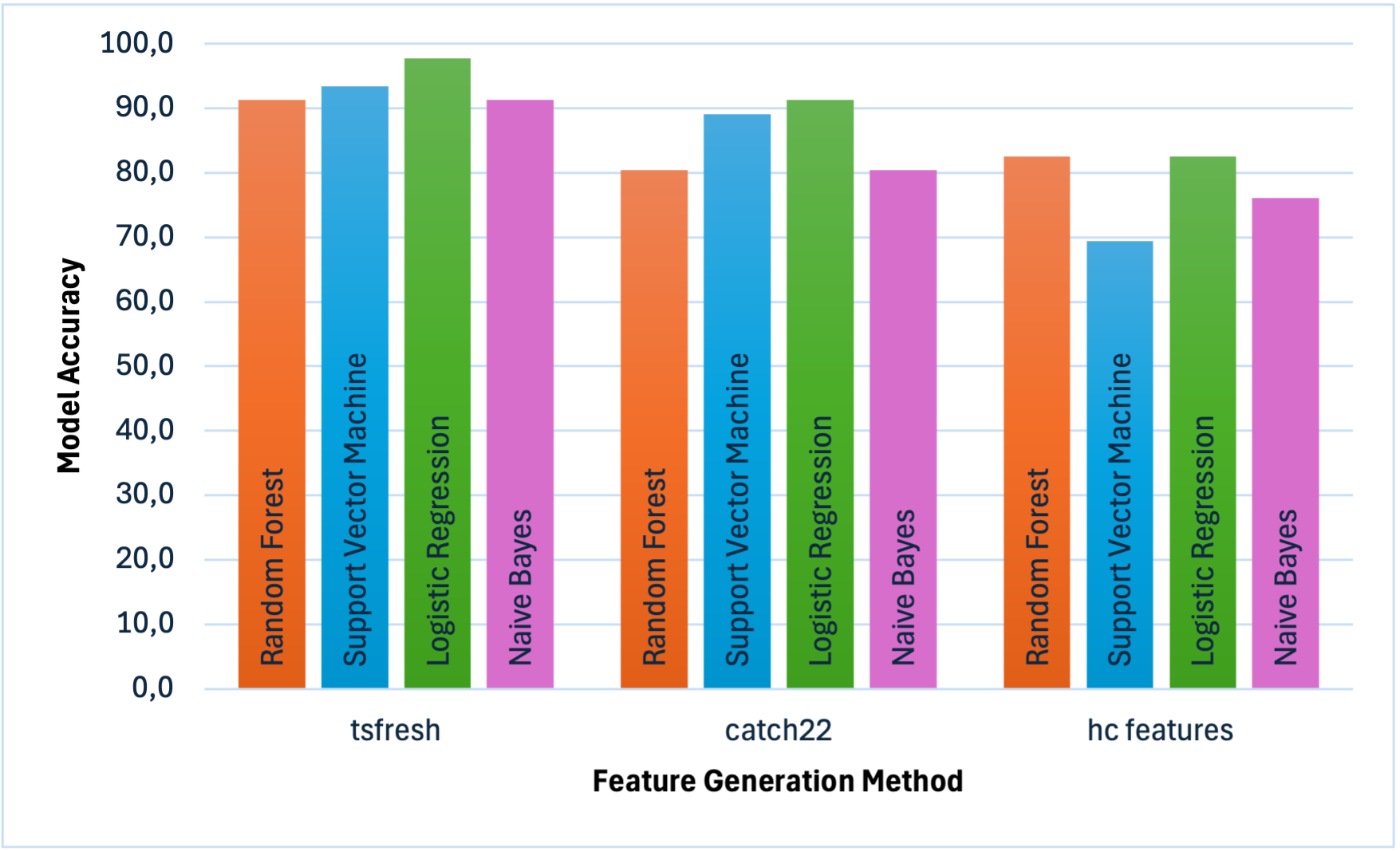}
\caption{Machine Learning Accuracy Results of each Model of the three Approaches} \label{Distribution}
\end{figure}

The evaluation of our approach consists of a detailed description of the dataset, followed by an explanation of the results from applying the described processing chain.\\
Of the N=46 participants in the experiment, 20 were women, 25 were men, and one unspecified, aged from 13 to 71. The sample consists of 5 pupils, 3 students, 23 employees, 4 civil servants, 5 self-employed individuals, 6 retirees, and 2 participants who did not specify their occupations, with the majority holding university degrees (23), followed by 10 with a high school diploma (Abitur).\\
The internal consistency of our user satisfaction scale, as measured by Cronbach's Alpha, is 0.78, which is considered an acceptable value and borders on being a good result according to \cite{george_spss_2003}. The participants reported a user satisfaction score of 3.4 on average, with a standard deviation of 0.81, a minimum of 1.4, and a maximum of 4.8. \\
The Duration of an interaction was 2 minutes and 12 seconds on average, with a standard deviation of 61 seconds, a minimum of 35 seconds, and a maximum duration of 5 minutes and 35 seconds. On average, the interactions comprised seven turns, with a minimum of 2, a maximum of 16, and a standard deviation of 3 turns. 
The resulting machine learning dataset consists of 15 instances with low user satisfaction ratings and 31 instances with medium to high user satisfaction ratings. The accuracy results of each of the three methods are depicted in Figure \ref{Distribution}, and the resulting metrics are listed in Table \ref{tab:my-table}.  
  
\begin{table}[]
\centering
\caption{Macro Averages of the Metrics of the Machine Learning Training for each of the three Feature Engineering Approaches.}
\label{tab:my-table}
\resizebox{\columnwidth}{!}{%
\begin{tabular}{@{}l|c|c|c|c|c|l@{}}
\cmidrule(lr){2-6}
                                                         & \textbf{Precision} & \textbf{Recall} & \textbf{F1-Score} & \textbf{Accuracy} & \textbf{ROC-AUC} &  \\ \cmidrule(r){1-6}
\multicolumn{1}{|l|}{tsfresh Random Forest}               & 0,92               & 0,88            & 0,90              & 91,30             & 0,96             &  \\ \cmidrule(r){1-6}
\multicolumn{1}{|l|}{tsfresh Support Vector Machine}     & 0,94               & 0,97            & 0,95              & 93,40             & 0,99             &  \\ \cmidrule(r){1-6}
\multicolumn{1}{|l|}{tsfresh Logistic Regression}        & 0,97               & 0,98            & 0,98              & 97,80             & 0,99             &  \\ \cmidrule(r){1-6}
\multicolumn{1}{|l|}{tsfresh Naive Bayes}                 & 0,90               & 0,90            & 0,90              & 91,30             & 0,92             &  \\ \cmidrule(r){1-6}
\multicolumn{1}{|l|}{catch22 Random Forest}               & 0,83               & 0,72            & 0,74              & 80,43             & 0,78             &  \\ \cmidrule(r){1-6}
\multicolumn{1}{|l|}{catch22 Support Vector Machine}     & 0,82               & 0,82            & 0,83              & 89,13             & 0,93             &  \\ \cmidrule(r){1-6}
\multicolumn{1}{|l|}{catch22 Logistic Regression}        & 0,92               & 0,88            & 0,90              & 91,30             & 0,96             &  \\ \cmidrule(r){1-6}
\multicolumn{1}{|l|}{catch22 Naive Bayes}                 & 0,78               & 0,77            & 0,77              & 80,43             & 0,83             &  \\ \cmidrule(r){1-6}
\multicolumn{1}{|l|}{hc features Random Forest}           & 0,81               & 0,78            & 0,79              & 82,60             & 0,78             &  \\ \cmidrule(r){1-6}
\multicolumn{1}{|l|}{hc features Support Vector Machine} & 0,86               & 0,60            & 0,59              & 69,50             & 0,80             &  \\ \cmidrule(r){1-6}
\multicolumn{1}{|l|}{hc features Logistic Regression}    & 0,82               & 0,77            & 0,79              & 82,60             & 0,84             &  \\ \cmidrule(r){1-6}
\multicolumn{1}{|l|}{hc features Naive Bayes}             & 0,76               & 0,67            & 0,74              & 76,10             & 0,79             &  \\ \cmidrule(r){1-6}
\end{tabular}
}
\end{table}
The classification results based on features computed using the tsfresh library achieved accuracies of 91.3\% for Random Forest and Naive Bayes, 93.4\% for Support Vector Machine, and 97.8\% for Logistic Regression. Precision, Recall, F1-Score, and ROC-AUC performance metrics were generally above 0.9, except in one instance. 
Out of 783 features per time series the 10 most important ones selected for training  (with respective significance levels), were: \\ twice distance (p<0.05 \& p<0.001), facial expression fear (p<0.001), width (p<0.001), head heading (p<0.001), vertical motion (p<0.001), head tilt (p<0.001), facial expression surprise (p<0.05), and head pitch (p<0.05).\\ The top ten features result from applying a Fast Fourier Transform to the social signal time series and include coefficients from the real, imaginary, and angle components. \\
For the method utilizing the catch22 library, the respective models achieved classification accuracies ranging between 80,4\% and 91,3\% with Logistic Regression. The performance metrics were below those of the tsfresh approach. Out of 506 calculated features the ten most relevant are out of the time series of, in brackets the feature category and significance,\\ trunk tilt (self-affine scaling, p<0.05), head heading (extreme event timing - positive outlier timing, p<0.05), vertical motion (nonlinear autocorrelation, p<0.05), width (symbolic, p<0.03), height (distribution shape, p<0.03), depth (linear autocorrelation structure, p<0.03), distance (nonlinear autocorrelation, p<0.03), distance (linear autocorrelation, p<0.03), facial expression disgust (self-affine scaling, p<0.03) and facial expression happiness (extreme event timing - positive outlier timing, p<0.03).\\ For further explanation, we refer to the catch22 feature description \cite{noauthor_feature_2024}. \\
The models trained using handcrafted feature methods achieved classification accuracies ranging from 69.5\% to 86.6\% with Random Forest and Logistic Regression. The performance metrics ranged from 0.59 to 0.86 within an acceptable range, though they exhibited more variability than the other approaches. Out of 766 automatic generated features the 10 best time series features were\\ trunk heading (Window 2 Frequency Zone 5, p<0.03), trunk pitch (longest period Zone 2, p>0.05), trunk pitch (shortest period Zone 2, p<0.03), head heading (Window 1 Frequency Zone 5,p<0.05), body speed (Count Zone Changes,p>0.05), body speed (shortest period Zone 10,p<0.05), body speed (Window 1 Frequency Zone 8,p<0.05), distance (Window 5 Frequency Zone3,p<0.03), facial expression disgust (Window 1 Frequency Zone 2,p>0.05), and facial expression neutral (Count Zone Changes,p<0.03).\\ For the complete feature tables we refer to the online Appendix \cite{michael_icsr2025}.
\section{Discussion}
In this study, we investigated the extent to which user satisfaction in human-robot interactions can be classified based on social signals without elaborately annotating the interaction data. To achieve realistic results, we used interaction data collected in a museum through video recordings and questionnaires.
Our approach includes video recording in the field, extraction of social signal metrics as time series, and feature generation using the tsfresh and catch22 libraries, as well as a hand-crafted approach \\
Due to the small number of N=46 instances, an adapted training procedure (LOOCV) was used to minimize the risk of overfitting on the small data set, the model results need to be interpreted with this limitation.
The tsfresh approach generates features with very high significance levels, based on which we achieve the best classification results in comparison. These results underline the strength of tsfresh in capturing detailed patterns in the data, as but also may indicate overfitting to some degree. 
With the catch22 approach, we also generate statistically significant features that allow us to classify user satisfaction with acceptable but lower performance metrics than tsfresh. One advantage of this approach is that we generate fewer features per time series, which makes processing more efficient while still achieving acceptable performance.
The hand-crafted features achieve solid classification results, but fall short of the other approaches. We hypothesize that while the hand-crafted features capture the dynamics of the time series, as evidenced by the features that reflect the frequency of the classified zones, they may not be nuanced enough to enable higher classification performance.\\
A clear limitation of the in-the-wild condition is that the necessary selection of interactions to compile a dataset resulted in a relatively small dataset. Another limitation is the additional necessary cameras and the privacy policy button to start the conversation, which can lead to influences in the users natural behavior. 
Common to all approaches is that the social signals were extracted from head and torso orientation and facial expressions, and are significant in all three approaches. These features contribute to classification performance through their temporal and distributional characteristics, likely reflecting the underlying affective state associated with the resulting user satisfaction.
The Fast Fourier Transform in the tsfresh approach captured statistically significant features based on the oscillatory properties of the underlying time series and reflected these in the form of coefficients. The use of catch22 additionally shows that self-affine scaling and extreme event timings indicate sudden changes in behavior or outlier events that could influence user satisfaction. The features of the hand-crafted approach complement the observations with more comprehensible features, such as the longest trunk pitch in the second zone, which corresponds to leaning forward. These observations are consistent with Tickle-Degnen and Rosenthal's \cite{tickle-degnen_nature_1990} findings about rapport, where certain gestures indicate rapport. The observations also demonstrate the need to examine further the relationship between oscillation characteristics, e.g., the relation of frequency and amplitude of head movements, and extreme events in time series of social signals and concepts like user satisfaction, as well as higher-level psychological concepts such as rapport. This can reveal the significance of specific gestures or movement patterns, especially if recognizable patterns exist among varying groups of user satisfaction. The feature engineering methods shown do not allow any direct statement about the significance of specific gestures or similar social cues. Still, they seem to map the complex relationships of social signals to higher-order constructs such as user satisfaction.\\
Although we did not test the transferability directly, we hypothesize that our approach of combining social signal metrics with time series classification might also be suitable for predicting other psychological constructs such as engagement, relationship, or other quality scales, due to the highly significant features and the already known relationship between certain body language and e.g. relationship and engagement.
Further research should focus on investigating the approach's generalizability, its transferability to other social interaction agents (SIA), and its practical applicability. To improve applicability, techniques should be explored that allow rapid adaptation to the specific use case and, for example, require only a small number of interactions and questionnaires, such as transfer learning models.
In this study, we examined a stationary robot and were able to take pictures of the users with additional cameras while they were standing relatively still. The transferability of the approach to other agents poses a challenge, as the users' body language depends on the interaction setting. This is particularly true for mobile, humanoid robots and digital agents, as the observation angle and interaction dynamics change here, creating challenges in extracting social signal features.

\section{Conclusion}
The demonstrated methods pave the way for predicting user satisfaction for automatic and semi-automatic improvements by identifying the user interactions that leave users unsatisfied.
Our results show that feature extraction based on social signal time series is extremely promising. In this paper, we demonstrated that it is possible to classify a user's satisfaction rating of the corresponding questionnaire using features automatically derived from time series of social signal metrics from body language, facial expressions, and distance.  
Based on a dataset of N=46 in-the-wild single-user interactions recorded from the SIA's point of view, we compared three automatic feature engineering approaches, selected features based on the statistical significance, and trained different machine learning models. Our results provide insight into the trade-off between relatively tractable features and high precision in performance metrics.
Despite the small data set, the approaches can provide highly relevant features for machine learning by achieving sufficient accuracies and performance metrics. 
The presented approach has the advantage of not relying on extensive data annotation, as existing approaches do, and instead uses the user's given user satisfaction rating after the interaction. 

\begin{credits}
\subsubsection{\ackname} We thank our collaboration partner DB Systel GmbH and the \\ Deutsches Museum Bonn, for their assistance and contributions. The Research activities were reviewed and approved by the Ethics Research Committee of TH Köln (application no. THK-2023-0004).

\subsubsection{\discintname}
The authors acknowledge the financial support by the Federal Ministry of Education and Research of Germany in the framework FH-Kooperativ 2-2019 (project number 13FH504KX9).
\end{credits}
%
%
%
%
\bibliographystyle{splncs04}
\bibliography{Literatur.bib} 
\end{document}